\documentstyle[fleqn,12pt]{article}
\input epsf
 
\def\bo{{\raise.15ex\hbox{\large$\Box$}}}

\def\dag{^{\dagger}{}}

\def\ordless{{\lower2mm\hbox{$\,\stackrel{\textstyle <}{\sim}\, $}}}
\def\ordmore{{\lower2mm\hbox{$\,\stackrel{\textstyle >}{\sim}\, $}}}

\newtoks\slashfraction
\slashfraction={.13}
\def\slash#1{\setbox0\hbox{$\, #1$}
\setbox0\hbox to \the\slashfraction\wd0{\hss \box0}/\box0}

\def\leftrightarrowfill{$\mathsurround=0pt \mathord\leftarrow \mkern-6mu
        \cleaders\hbox{$\mkern-2mu \mathord- \mkern-2mu$}\hfill
        \mkern-6mu \mathord\rightarrow$}
\def\overleftrightarrow#1{\vbox{\ialign{##\crcr
        \leftrightarrowfill\crcr\noalign{\kern-1pt\nointerlineskip}
        $\hfil\displaystyle{#1}\hfil$\crcr}}}
\def\startarray{\left( \begin{array}}
\def\finarray{\end{array} \right)}
\def\starteq{
\begin{eqnarray}}
\def\fineq{\end{eqnarray}
}

\catcode`@=11
\def\underline#1{\relax\ifmmode\@@underline#1\else
$\@@underline{\hbox{#1}}$\relax\fi}
\catcode`@=12
 
\newskip\humongous \humongous=0pt plus 1000pt minus 1000pt

\newif\ifdtup
 
\def\textcite#1{Ref.~{\cite{#1}}}

\def\next{\nonumber \\}
\def\thefootnote{\fnsymbol{footnote}}
 
\def\author#1#2{{\bf #1} \\ {\em #2}\vspace{5mm}}

\def\bold#1{\setbox0=\hbox{$#1$}%
     \kern-.025em\copy0\kern-\wd0
     \kern.05em\copy0\kern-\wd0
     \kern-.025em\raise.0433em\box0 }

\def\title#1#2#3#4#5{\thispagestyle{empty}
        \begin{center} \vspace*{1cm} { \bf #3} \\[.5in] {#4{}}
        \end{center} \vfill \centerline{ ABSTRACT}
   {\nopagebreak \noindent\begin{quotation}\noindent {\small #5}
   \end{quotation}} \vfill {#2} \hfill\begin{tabular}{r} {#1} 
        \end{tabular}  \newpage
        \def\thefootnote{\arabic{footnote}}}
\def\prefer{\section*{}
    \list{[\arabic{enumi}]}{\usecounter{enumi}\settowidth\labelwidth{[000]}
      \leftmargin\labelwidth\advance\leftmargin\labelsep \rightmargin=0pt}
        \small \sfcode`\.=1000\relax}

\def\refer#1{\section*{\large \sc {#1}}
    \list{\arabic{enumi}.}{\usecounter{enumi}\settowidth\labelwidth{[000]}
      \leftmargin\labelwidth\advance\leftmargin\labelsep \rightmargin=0pt}
        \raggedright \small \sfcode`\.=1000\relax}

\def\ReFer#1#2{\section*{\large\sc#1}
    \list{[\arabic{enumi}]}{\usecounter{enumi}\settowidth\labelwidth{#2}
      \leftmargin\labelwidth\advance\leftmargin\labelsep \rightmargin=0pt}
        \raggedright \small \sfcode`\.=1000\relax}

\def\REFER#1#2{\section*{\large\sc#1}
    \list{#2 {enumi}.}{\usecounter{enumi}\settowidth\labelwidth{[000]}
      \leftmargin\labelwidth\advance\leftmargin\labelsep \rightmargin=0pt}
        \raggedright \small \sfcode`\.=1000\relax}

\def\startbib{\vspace{1in}\begin{refer}{References}
\small\frenchspacing\nopagebreak}
\def\endbib{\end{refer} \normalsize \nonfrenchspacing}
\def\startfig{\newpage \centerline{{\sl Figure captions}} \begin{itemize}}

\def\endfig{\end{itemize}}
 
\newcommand{\be}{\begin{equation}}
\newcommand{\ee}{\end{equation}}
\newcommand{\bea}{\begin{eqnarray}}
\newcommand{\eea}{\end{eqnarray}}

\newcommand{\AmS}{{\protect\the\textfont2
  A\kern-.1667em\lower.5ex\hbox{M}\kern-.125emS}} 
\begin{document}
\title{September 2000} {PC084.0900}
{ $B \rightarrow K\pi$ DECAYS  AND THE  WEAK PHASE ANGLE $\gamma\dag$  
\footnotetext{${\dag}$  
Talk given at the QCD Euroconference 00, Montpellier 6-12 July 2000}}
{\author{T. N. Pham} {Centre de Physique Th\'eorique, \\
Centre National de la Recherche Scientifique, UMR 7644, \\  
Ecole Polytechnique, 91128 Palaiseau Cedex, France}}   
{The large 
branching ratios  for  $B \rightarrow K\pi$ decays as observed  by the CLEO
Collaboration  indicate that penguin interactions
contribute a major part to the decay rates and provide an interference
between the Cabibbo-suppressed tree and penguin contributions resulting in
a CP-asymmetry between the $B \rightarrow K\pi$ and its charge conjugate
mode. The CP-averaged decay rates  depend also on  the weak phase
$\gamma$ and give us a  determination of this phase. In this talk, I
would like to report on a recent analysis of $B \rightarrow K\pi$ decays
using factorisation model with final state interaction 
phase shift included. We find that factorisation
seems to describe qualitatively the latest CLEO data. We also obtain   
 a relation for the branching
ratios independent of the strength of the strong penguin interactions.
This relation gives a central value of $0.60 \times 10^{-5}$ for 
${\mathcal B}(\bar{B}^{0} \rightarrow \bar{K}^{0}\pi^{0})$, somewhat
smaller than the latest CLEO measurement. 
We also find that a ratio obtained from the 
CP-averaged $B \rightarrow K\pi$ decay rates could be used
to test the factorisation model and to determine the weak angle $\gamma$
with more precise data, though the latest CLEO data seem to favor
$\gamma$ in the range  $90^{\circ}-120^{\circ}$.}

With the measurement of all the four $B \rightarrow K\pi$ branching ratios, 
we seem to have a qualitative understanding of the $B \rightarrow K\pi$ decays. 
The  measured CP-averaged branching ratios(${\mathcal B}$)   by 
the CLEO Collaboration \cite{cleo2} show  that the penguin interactions
dominates the $B \rightarrow K\pi$ decays, as predicted by factorisation. 
The strong penguin
amplitude, because of the large CKM factors, becomes much larger than the
tree-level terms which are Cabibbo-suppressed and the non-leptonic
interaction for $B \rightarrow K\pi$ is dominated by an $ I = 1/2 $ amplitude.
This is borne out by the CLEO data which give
$ {\mathcal B}(B^{-} \rightarrow \bar{K}^{0}\pi^{-}\simeq 2{\mathcal B}(B^{-} \rightarrow
K^{-}\pi^{0}) $ and 
${\mathcal B}(B^{-} \rightarrow \bar{K}^{0}\pi^{-}) \simeq {\mathcal B}(\bar{B}^{0}
\rightarrow K^{-}\pi^{+})$~:
\begin{eqnarray}  
{\mathcal B} (B^{+} \rightarrow K^{+} \pi^0)
&=&(11.6^{+3.0+1.4}_{-2.7-1.3}) \times 10^{-6},  \nonumber  \\
{\mathcal B} (B^{+} \rightarrow K^0 \pi^{+})  &=& 
(18.2^{+4.6}_{-4.0}\pm 1.6) \times 10^{-6},   \nonumber  \\
{\mathcal B} (B^0\rightarrow K^{+} \pi^{-})  &=& 
(17.2^{+2.5}_{-2.4}\pm 1.2) \times 10^{-6}, \nonumber \\ 
{\mathcal B} (B^0  \rightarrow K^0\pi^0)\, \  &=& \
(14.6^{+5.9+2.4}_{-5.1-3.3}) \times 10^{-6}.
\end{eqnarray}
If the strength of the  interference between the tree-level and penguin
contributions is known, a determination of the weak phase $\gamma$ 
could be done in principle. Previous works \cite{Deshpande,Ali} shows that
factorisation model produces sufficient $B \rightarrow K\pi$ decay rates, in   
qualitative agreement with the CLEO measured values. Also, as argued  
in \cite{Neubert2}, for these very 
energetic decays,  because of color transparency, factorisation should be 
a good approximation for $B \rightarrow K\pi$ decays if the Wilson coefficients 
are evaluated at a scale $\mu =O(m_{b})$. Infact, recent hard scattering
calculations with perturbative QCD shows that factorisation is valid
up to corrections of the order $\Lambda_{\rm QCD}/m_{b}$ \cite{Beneke}. 
It is thus 
encouraging to use factorisation to analyse the $B \rightarrow K\pi$ decays,
bearing in mind that there are important 
theoretical uncertainties in the long-distance hadronic matrix elements, 
as the heavy to light form factors for the vector current and the value of
the current 
$s$ quark mass are currently not determined with good accuracy. 
In this talk, I
would like to report on a recent work \cite{Isola} on the $B \rightarrow K\pi$ 
decays as a possible way to measure the angle $\gamma$ and to see direct 
CP violation. 

In the standard model, the effective Hamiltonian for $B \rightarrow K\pi$ decays
are given by \cite{Buras,Ciuchini,Fleischer1,Kramer,Deshpande1},
\bea
\kern -0.7cm H_{\rm eff} = {G_F\over \sqrt{2}}[V_{ub}V^*_{us}(
c_1O_{1}^{u} + c_2O_{2}^{u})  + V_{cb}V^*_{cs}(c_1O_{1}^{c} 
 + \  c_2 O_{2}^{c}) -\sum_{i=3}^{10}V_{tb}V^*_{ts} c_iO_i]  
+ {\rm h.c.}\,.
\label{hw}
\eea
in standard notation.
At next-to-leading logarithms, $c_{i}$ take the form of an effective
Wilson coefficients $c_{i}^{\rm eff}$ which  contain also the penguin 
contribution from the $c$ quark loop and 
are given in \cite{Fleischer1,Deshpande1}.

The parameters  $V_{ub}$ etc. are the flavor-~changing charged current 
couplings of the weak gauge boson $W^{\pm}$ with the quarks as given by 
the Cabibbo-Kobayashi-Maskawa (CKM) quark mixing matrix 
$V$. $V$ is usually defined as the unitary transformation
relating the the weak interaction eigenstate of quarks to their mass 
eigenstate \cite{PDG}:
\be
\pmatrix{d' \cr s' \cr b'}= \pmatrix{V_{ud}& V_{us} & V_{ub} \cr
V_{cd} & V_{cs} & V_{cb} \cr 
V_{td} & V_{ts} & V_{tb}} \pmatrix{d \cr s \cr
b}
\label{Vckm} 
\ee
where $d,s,b$ and $d',s',b'$ are respectively the  mass eigenstates
and weak interaction eigenstates for the charge $Q=-1/3$ quarks. 
Since the neutral current is not affected by the unitary transformation
on the quark fields, 
flavor-changing neutral current is absent at the tree-level as implied
by the GIM mechanism. The unitarity condition $VV^{\dag}=1$ 
gives, for the (db) elements relevant to $B$ decays \cite{PDG}
\be
V_{ud}V_{ub}^{*} + V_{cd}V_{cb}^{*} + V_{td}V_{tb}^{*} =0 
\label{db}  
\ee
This can be represented by a triangle \cite{PDG} with the three 
angles $\alpha$, $\beta$
and $\gamma$ expressed in terms of the CKM matrix elements as \cite{Nir}:
\bea
&& \alpha = arg(-V_{td}V_{tb}^{*}/V_{ud}V_{ub}^{*})  \nonumber\\
&&  \beta = arg(-V_{cd}V_{cb}^{*}/V_{td}V_{tb}^{*})  \nonumber\\
&&  \gamma = arg(-V_{ud}V_{ub}^{*}/V_{cd}V_{cb}^{*})
\label{arg}
\eea
The angle $\gamma$  enters the $B \rightarrow K\pi$ decay amplitudes  through
the factor 
$V_{ub}V_{us}^*/V_{tb}V_{ts}^* $ which can be approximated by
$-(|V_{ub}|/|V_{cb}|)\times (|V_{cd}|/|V_{ud}|)\exp(-i\,\gamma)$
after neglecting terms of the order $O(\lambda^{5})$ in the (bs)
unitarity triangle, $\lambda $ being the Cabibbo angle in the 
Wolfenstein parametrisation of the CKM quark mixing matrix. The $B \rightarrow
K\pi$ decay amplitudes, expressed in terms of the $I=1/2$ and $I=3/2$
isospin amplitudes are given by \cite{Deshpande},

\bea
A_{K^-\pi^0} &=& 
 {2\over 3} B_3e^{i\delta_3} + \sqrt{{1\over 3}} 
(A_1+B_1)e^{i\delta_1}, \ \ \  \nonumber\\
A_{\bar K^0\pi^-} &=&
{\sqrt{2}\over 3} B_3e^{i\delta_3} - \sqrt{{2\over 3}} 
(A_1+B_1)e^{i\delta_1},\nonumber\\
A_{K^-\pi^+} &=&
{\sqrt{2}\over 3} B_3e^{i\delta_3} + \sqrt{{2\over 3}} 
(A_1-B_1)e^{i\delta_1},\ \ \ \nonumber\\
A_{\bar K^0\pi^0} &=& 
{2\over 3} B_3e^{i\delta_3} - \sqrt{{1\over 3}} 
(A_1-B_1)e^{i\delta_1},
\label{amplitude}
\eea
$A_{1}$ is the sum of  the strong penguin $A_{1}^{\rm S} $
and the $I=0$  tree-level $A_{1}^{\rm T}$ as well as the $I=0$ 
electroweak penguin $A_{1}^{\rm W}$ contributions 
to the $B \rightarrow K\pi$ $I=1/2$ amplitude;  similarly  
$ B_{1}$ is the sum of the
$I=1$ tree-level $B^{T}_{1}$ and electroweak penguin $B^{W}_{1}$
contribution to the $I=1/2$ amplitude, and $B_{3}$ is the sum of the
$I=1$  tree-level $B^{T}_{3}$ and electroweak penguin $B^{W}_{3}$
contribution to the $I=3/2$ amplitude. $\delta_1 $ and $ \delta_3$ are,
respectively, the  elastic  $\pi K \rightarrow \pi K$ $I=1/2$ and $I=3/2$ final 
state interaction (FSI) phase shift at the $B$ mass. 
The inelastic FSI contributions 
are also included through the internal quark loop contributions to the
penguin operators, for which the Wilson coefficients now have 
an absorptive  part and are given in \cite{Fleischer1,Deshpande1,Hou}.
The $B \rightarrow K\pi$ isospin amplitudes in the factorisation model 
are given by \cite{Deshpande},
\bea
&&  A_1^{\rm T} = i{\sqrt{3}\over 4}\, V_{ub}V_{us}^*\, r\, 
a_2,\nonumber\\
&& B_1^{\rm T} = i {1\over 2\sqrt{3}}\, V_{ub}V_{us}^*\, r 
\left[-{1\over 2}a_2 
+a_1 X\right],\nonumber\\
&& B_3^{\rm T} = i{1\over 2}\, V_{ub}V_{us}^*\, r 
\left[a_2 + a_1 X\right],
\nonumber\\
&& A_1^{\rm S} = -i{\sqrt{3}\over 2}\, V_{tb}V_{ts}^* r
\left[ a_4 + a_6 Y\right],
  \ B_1^{\rm S} = B_3^{\rm S} = 0,\nonumber\\
&& A_1^{\rm W} = -i{\sqrt{3}\over 8}\, V_{tb}V_{ts}^*r
\left[a_8Y +  a_{10}\right],
\nonumber\\
&& B_1^{\rm W} = i{\sqrt{3}\over 4}\, V_{tb}V_{ts}^*r
\left[{1\over 2}a_8Y 
+ {1\over 2}a_{10} 
+\left(a_7 - a_9\right)X\right],
\nonumber\\
&& B_3^{\rm W} = - i{3\over 4}\, V_{tb}V_{ts}^* r
\left[ a_8Y + a_{10} 
- (a_7 - a_9)X\right] 
\eea
where $r = G_F\, f_K F^{B\pi}_0(m^2_K) (m_B^2-m_\pi^2)$ ,\qquad
$X= (f_\pi/f_K)[F^{BK}_0(m^2_\pi)/F^{B\pi}_0(m^2_K)] 
[(m_B^2-m^2_K)/(m_B^2-m_\pi^2)]$,
$Y = 2m^2_K/[(m_s+ m_q)(m_b-m_q)]$ with $q=u,\ d$ for 
$\pi^{\pm,0}$ final states, respectively. In this analysis, 
$f_{\pi} =133 \,\rm MeV$, $f_{K} =158\,\rm MeV$, 
$ F_{0}^{B\pi}(0) = 0.33$, $ F_{0}^{BK}(0) = 0.38$ \cite{Ali,Bauer}~;
$|V_{cb}|=0.0395, |V_{cd}|=0.224 $ and $|V_{ub}|/|V_{cb}|=0.08$ \cite{PDG}.
The value of $m_{s}$ is not known to a good accuracy, but a value 
around $(100-120\rm )\, MeV$
inferred from $m_{K^{*}} - m_{\rho}$, $m_{D_{s}^{+}}- m_{D^{+}}$ 
and $m_{B_{s}^{0}}- m_{B^{0}}$ mass differences
\cite{Colangelo} seems not unreasonable and in this work, 
we use $m_{s} =120\,\rm MeV$.
$a_{j}$ are  the effective Wilson coefficients after Fierz reordering
in factorisation model and are given by \cite{Isola}
\bea
&& a_1 = \ 0.07    , \ \ \ a_2=  \ 1.05  , \nonumber\\
&& a_4 = -0.043 - 0.016i , \ \ a_6= -0.054 - 0.016i  ,
\label{coeff}
\eea
for the contributions from the
tree-level and  the strong penguin operators at $N_c =3$ and 
$m_{b} = 5.0\, \rm GeV$. 
The strong penguin contribution   $P = a_4 + a_6 Y$, 
as obtained from Eq.(\ref{coeff}) is enhanced by the
charm quark loop which increases the amplitude by $30\%$ as pointed out 
in \cite{Fleischer1}. This enhancement brings the predicted branching
ratios closer to the CLEO measured values, as shown in Fig.1. where
the CP-averaged $B \rightarrow K\pi$ branching ratios obtained for 
$\gamma = 70^\circ$ \cite{Ali}, are plotted against
the rescattering phase difference $\delta = \delta_{3} - \delta_{1}$. 
\begin{figure}[hbp]
\centering
\leavevmode
\epsfxsize=4.8cm
\epsffile{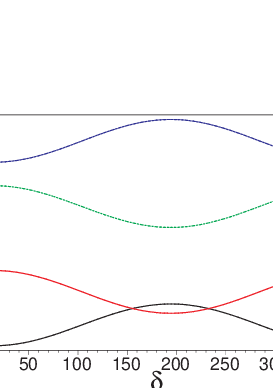}
\caption{${\mathcal B}(B\rightarrow K\pi)$ vs. $\delta$ for $\gamma = 70^\circ$.
The curves  
(a), (b), (c), (d)  are for the CP-averaged branching ratios
$B^-\rightarrow K^-\pi^0,\ \bar K^0 \pi^-$ and 
$\bar B^0 \rightarrow K^-\pi^+,\ \bar K^0\pi^0$, respectively.}
\label{fig:BR}
\end{figure}

For a determination of $\gamma$, two 
quantities obtained from the sum
of the two CP-averaged decay rates 
 $\Gamma_{B^-} = \Gamma(B^{-} \rightarrow K^- {\pi}^0) + 
\Gamma(B^{-} \rightarrow \bar{K}^{0} {\pi}^{-})$ and 
$\Gamma_{B^0} = \Gamma(\bar{B}^{0} \rightarrow K^- {\pi}^+) + 
\Gamma(\bar{B}^{0} \rightarrow \bar{K}^{0} {\pi}^{0})$ which are independent of
$\delta$ could be used \cite{Isola}.
As the CP-averaged $B \rightarrow K\pi$ decay rates depend  on $\gamma$,
the  computed partial rates $\Gamma_{B^{-}} $ 
and $\Gamma_{B^0} $ would now lie
between the upper and lower limit corresponding to $\cos(\gamma) =1 $ 
and $\cos(\gamma) =-1 $, respectively.
As shown in Fig.2, where the corresponding CP-averaged branching
ratios (${\mathcal B}_{B^0} $ and
${\mathcal B}_{B^-}$)
for $\Gamma_{B^{-}} $ and $\Gamma_{\bar{B}^{0}} $ are plotted against 
$\gamma$, the factorisation model values with the 
BWS form factors \cite{Bauer}
seem somewhat smaller
than the CLEO central values by about $10-20\% $. Also,  
${\mathcal B}_{B^{-}} > {\mathcal B}_{\bar{B}^{0}} $
while the data give
${\mathcal B}_{B^{-}} < {\mathcal B}_{\bar{B}^{0}} $
by a small amount which could be due to a large
measured $\bar{B}^{0} \rightarrow \bar{K}^{0}\pi^{0} $ branching ratio.    
\begin{figure}[hbp]
\centering
\leavevmode
\epsfxsize=4.8cm
\epsffile{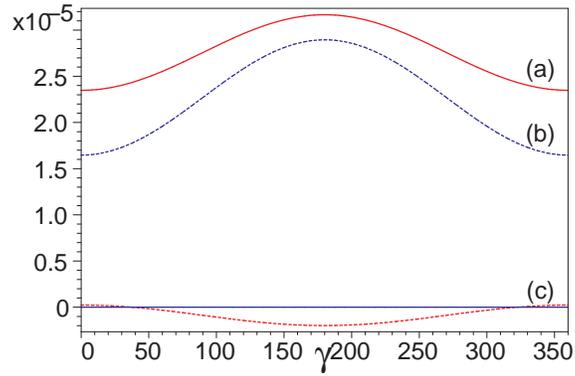}
\caption{${\mathcal B}_{B^-}$ (a), ${\mathcal B}_{\bar{B}^0}$ (b), $\Delta$ (c) vs. $\gamma$}
\label{fig:facttest}
\end{figure}

Note that smaller values for the form factors 
could easily accommodate the latest CLEO measured values, if a smaller
value for $m_{s}$, e.g, in the range $(80-100)\,\rm MeV$ is used. What
one learns from this analysis is that $B \rightarrow K\pi$ decays are 
penguin-dominated and the strength of the penguin interactions as obtained by
perturbative QCD, produce sufficient $B \rightarrow K\pi$ decay rates and that
factorisation seems to work with an accuracy better than a factor of 2,
considering large uncertainties from the form factors and possible
non-factorisation terms inherent in the factorisation model
and the uncertainties in the penguin amplitude which is sensitive to the
current $s$ quark mass. Since  the four $B \rightarrow K\pi$ decay rates 
depend on only three amplitudes $A_{1} $, $B_{1} $ and $B_{3} $, it is
possible to derive a relation
between the decay rates independent of $A_{1}$. Thus, the
quantity $\Delta $ obtained from the decay rates,
\bea
&& \kern -0.7cm \Delta =\left\{ \Gamma(B^{-} \rightarrow \bar{K}^{0} {\pi}^{-})+\Gamma(\bar{B}^{0}
\rightarrow K^- {\pi}^+) \right. \nonumber\\
&& \kern -0.7cm  \ \  \ \ \ \ -  \left. 2\left[\Gamma(B^{-} \rightarrow K^- {\pi}^0)
+\Gamma(\bar{B}^{0} \rightarrow
     \bar{K}^{0} {\pi}^{0})\right]\right\}\tau_{B^0}\nonumber\\
&&\kern -0.7cm \ \ \ =  \left[-{4\over 3}|B_3|^2-{8\over {\sqrt 3}}{\rm Re} (B_3^* B_1 {\rm  e}^{i\delta})\right](C\tau_{B^0}) 
\label{delta}
\eea
is independent of  the strong penguin term. It is given by the
tree-level and electroweak penguin contributions. As can be seen 
from Fig.2, where its values for  $\delta =0$ 
are plotted against $\gamma $. $\Delta$ is of the order $O(10^{-6})$
compared with  ${\mathcal B}_{B^{-}} $ and ${\mathcal
  B}_{\bar{B}^{0}} $ which 
are in the range  $(1.6-3.0)\times 10^{-5}$. Thus, to this
level of accuracy, we can put $\Delta \simeq 0$ and obtain the relation
($r_b=\tau_{\bar{B}^0}/ \tau_{B^-}$).
\be
{ r_b}{\mathcal B}_{\bar{K}^{0}
    {\pi}^{-}}+{\mathcal B}_{K^- {\pi}^+} = 
2\,\left[{\mathcal B}_{\bar{K}^{0} {\pi}^{0}}+
{ r_b}{\mathcal B}_{ K^- {\pi}^0}\right]\,.
\label{kpi0}
\ee
which can be used to test factorisation or to predict 
${\mathcal B}(\bar B^0\rightarrow \bar K^0 {\pi}^0)$ in terms of the 
other measured branching ratios. Eq.(\ref{kpi0}) then predicts a central value 
${\mathcal B}(\bar{B}^{0} \rightarrow\bar{K}^{0} 
{\pi}^{0})= 0.60\times 10^{-5}$.

Since  ${\mathcal B}_{\bar{K}^{0} {\pi}^{0}} $ is not known with good
accuracy at the moment, it is useful to use another quantity, defined as
\be
  { r_b}{\mathcal B}_{\bar{K}^{0}
    {\pi}^{-}}+{\mathcal B}_{K^- {\pi}^+} = (C_1\tau_{\bar{B}^{0}})
  \left[{1\over 3} |B_3|^2 +(|A_1|^2 + 
|B_1|^2)-{2\over \sqrt 3}{\rm Re}(B_3^* B_1 {e}^{i\delta})\right]
\label{B23}
\ee
which contains a negligible $\delta$-dependent term of the order 
$O(10^{-7})$. The quantity $R$ defined as
\be
R={{{\mathcal B}(B^-\rightarrow K^{-} {\pi}^{0})
+{\mathcal B}( B^-\rightarrow \bar{K}^{0} {\pi}^{-})}\over{
{\mathcal B}(B^-\rightarrow\bar{K}^{0}
    {\pi}^{-})+{{\mathcal B}(\bar B^0\rightarrow K^- {\pi}^+)}/{r_b}}}\,\, .
\label{R}
\ee
is thus essentially independent of $\delta$ and could also be used to
obtain $\gamma$, as it does not suffer from large uncertainties in the
form factors and in the CKM parameters. As can be seen in Fig.3, 
it is not possible to deduce a value for $\gamma$ with the present
data which give $R=(0.80\pm 0.25) $ 
as the  prediction for $R$ lies within
the experimental errors. If we could reduce the experimental
uncertainties to a level of less than $10\%$, we might be able to 
give a value for $\gamma$. Thus it is important to measure 
$B\rightarrow K\pi$ branching ratios to a high precision. Also shown in
Fig.3 are two other quantities more sensitive to $\gamma$, but involved 
${\mathcal B}(\bar B^0\rightarrow \bar K^0 {\pi}^0) $ and are given as \cite{Isola} 
\be
R_{1} = {\Gamma_{B^{-}} \over \Gamma_{\bar B^0}} \, \ ,\
R_{2} = {\Gamma_{B^{-}} \over (\Gamma_{B^{-}} +\Gamma_{\bar B^0})}
\label{ratios}
\ee
\begin{figure}[hbp]
\centering
\leavevmode
\epsfxsize=4.8cm
\epsffile{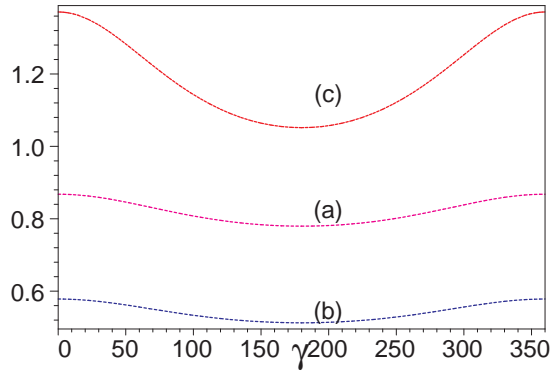}
\caption{The curves (a), (b), (c) are for $R$, $R2$, $R1$ respectively.}
\label{fig:R}
\end{figure}

Thus a better way to obtain $\gamma$ would be to use $R_{1}$ when a 
precise value for ${\mathcal B}(\bar B^0\rightarrow \bar K^0 {\pi}^0) $ will be
available. The  central
value of $0.80$ for $R$ corresponds to $\gamma = 110^{\circ}$, close to 
the   value  of $(113^{+25}_{-23})^{\circ}$ found
by the CLEO Collaboration in an analysis of all known charmless two-body
$B$ decays with the factorisation model \cite{cleo1}. It seems that 
the CLEO data favor a large  
$\gamma$ in the range $90^{\circ}-120^{\circ}$. With a large $\gamma$, 
for example, with the central value of $110^{\circ} $, as shown in Fig.4,
the predicted $B \rightarrow K\pi$ branching ratios are larger than that for
$\gamma = 70^{\circ}$ and are closer to the data. The data also show that
 $B^-\rightarrow \bar K^0 \pi^-$ and $\bar B^0 \rightarrow K^-\pi^+$ 
are the two largest modes with near-equal
branching ratios in qualitative agreement with factorisation. However,
for $\gamma = 70^{\circ}$, 
Fig.1 shows that these two largest branching ratios are quite apart,
except for $\delta < 50^{\circ}$ while for $\gamma = 110^{\circ}$,
Fig.4 suggests that these two
branching ratios are closer to each other only for
$\delta $ in the range $40^{\circ}-70^{\circ}$. 
With  $\gamma < 110^{\circ}$ and some adjustment of form 
factors, the current $s$ quark mass and CKM parameters, 
it might be possible to accommodate these two largest
branching ratios with  $\delta < 50^{\circ}$.

The  CP-asymmetries, as shown in \cite{Isola}, for $\gamma = 110^\circ$,    
are in the range $\pm(0.04)$ to $\pm(0.3)$
for the preferred values of $\delta$ in the range
$(40-70)^{\circ}$, but could be smaller for $\delta < 50^{\circ}$. 
The CLEO measurements \cite{cleo3} however, do not show any 
large CP-asymmetry in $B \rightarrow K\pi$ decays, but the errors are still
too large to draw any conclusion at the moment. 

In conclusion, factorisation with enhancement of the strong penguin
contribution  seems to describe qualitatively
the $B \rightarrow K\pi$ decays, 
Further measurements will allow 
a more precise test of factorisation and a determination of
the weak  angle $\gamma $ from the FSI phase-independent
relations shown above.
\vspace{-1cm}
\begin{figure}[hbp]
\centering
\leavevmode
\epsfxsize=4.8cm
\epsffile{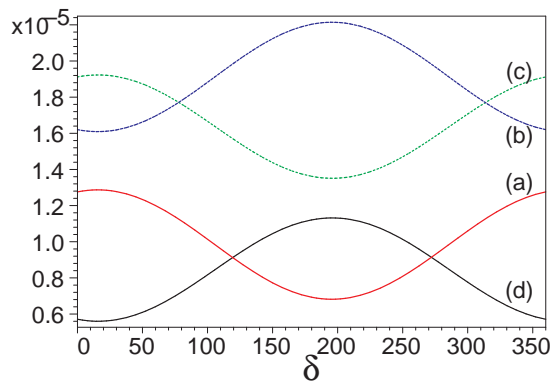}
\vspace{-1cm}
\caption{${\mathcal B}(B\rightarrow K\pi)$ vs. $\delta$ for $\gamma = 110^\circ$.
The curves  
(a), (b), (c), (d)  are for the CP-averaged branching ratios
$B^-\rightarrow K^-\pi^0,\ \bar K^0 \pi^-$ and 
$\bar B^0 \rightarrow K^-\pi^+,\ \bar K^0\pi^0$, respectively.}
\label{fig:BRc}
\end{figure}

I would like to thank S. Narison and the organisers of QCD00 for the
warm hospitality extended to me at Montpellier.

\end{document}